\documentclass[twocolumn,pre]{revtex4}

\usepackage{graphicx}
\usepackage{dcolumn}
\usepackage{bm}
\usepackage{graphics}
\usepackage{mathrsfs}

\newcommand{\be}{\begin{equation}}
\newcommand{\ee}{\end{equation}}
\newcommand{\bea}{\begin{eqnarray}}
\newcommand{\eea}{\end{eqnarray}}

\begin{document}

\title{Kinetics of viral self-assembly: the role of ss RNA antenna}

\author{Tao Hu}
\author{B. I. Shklovskii}
\affiliation{Theoretical Physics Institute, University of
Minnesota, Minneapolis, Minnesota 55455}

\date{\today}

\begin{abstract}
A big class of viruses self-assemble from a large number of
identical capsid proteins with long flexible N-terminal tails and
ss RNA. We study the role of the strong Coulomb interaction of
positive N-terminal tails with ss RNA in the kinetics of the in
vitro virus self-assembly. Capsid proteins stick to unassembled
chain of ss RNA (which we call "antenna") and slide on it towards
the assembly site. We show that at excess of capsid proteins such
one-dimensional diffusion accelerates self-assembly more than ten
times. On the other hand at excess of ss RNA, antenna slows
self-assembly down. Several experiments are proposed to verify the
role of ss RNA antenna.
\end{abstract}

\maketitle

\begin{figure}[htb]
\centering
\includegraphics[width=0.45 \textwidth]{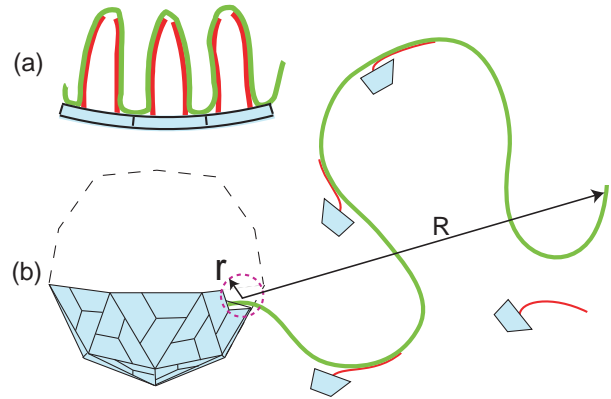}
\caption{(color online) (a) A blowup view from the inside of the
virus. The brush of positive N-terminal tails (red/dark gray line) is rooted at
the inner surface of the capsid (blue/light gray block). The ss RNA (green/gray line) strongly
interacts with the tails and glues all the CPs together. (b)
Schematic model of the capsid self-assembly. The unassembled ss
RNA makes an antenna of size $R$ for the one-dimensional
pathway of the CPs towards the capsid assembly site at the
capsid fragment (dashed circle with radius $r$ of the size of a CP.}
\label{fig:assembly}
\end{figure}

Viruses self-assemble in host cells from identical capsid proteins
(CPs) and their genome which in many cases is a long single stranded
(ss) RNA. Icosahedral viruses are formed from 60T CPs for only
certain triangulation number T such as 1, 3, 4, or 7,
etc~\cite{Book}. The Coulomb interaction between CP and ss RNA plays
an important role in their
self-assembly~\cite{Brancroft,adolph,Bruinsma1,Bruinsma}. Two recent
papers~\cite{Belyi,Hu} emphasized that CPs of a big class of T = 3 and
T= 4 viruses have long flexible N-terminal tails. They explored
the role played in the energetics of the virus structure by the
Coulomb interaction between the brush of positive N-terminal tails
rooted at the inner surface of the capsid and the negative ss RNA
molecule (see Fig. \ref{fig:assembly}a). It was shown~\cite{Hu} that
virus particles are most stable when the total length of ss RNA is
close to the total length of the tails. For such a structure the
absolute value of the total (negative) charge of ss RNA is
approximately two times larger than the charge of the capsid. This
conclusion agrees with available structural data. (Similar result
was obtained earlier~\cite{Bruinsma1} assuming that the positive
charge of CP is smeared on the inner surface of the capsid).

In this paper we continue to deal with electrostatic interaction
of N-terminal tails and ss RNA, but switch our attention from the
thermodynamics to the kinetics of in vitro self-assembly. Most of
papers on in vitro kinetics study self-assembly of an empty capsid
at much higher than biological concentrations of salt, where the
Coulomb repulsion of capsid proteins is screened and hydrophobic
interactions dominate~\cite{Zlotnick00,Casini}. In
Ref.~\cite{Casini} one can clearly discriminate the initial
nucleation "lag phase", followed by the "growth phase", where the
average mass of the assembled particles linearly grows with time.
The recent study of the kinetics of self-assembly with ss RNA
genome emphasizes that CPs stick to ss RNA before the
assembly~\cite{Vogt95,Zlotnick04}, so that a virus is assembled
actually from the linear CP-RNA complex. Not much is known about
the nucleation and growth phases of such assembly.

The goal of this paper is to understand the role of the large
length of ss RNA in kinetics of self-assembly at biological salt
concentrations. We assume that after nucleation (for example, at
one end of ss RNA) the capsid growth is limited by CP diffusion.
We calculate the acceleration of self-assembly, which originates from
the fact that due to the Coulomb interaction of N-terminal tails
with ss RNA, CPs stick to ss RNA and slide on it to the assembly
site. In this case, ss RNA plays the role of a large antenna
capturing CPs from the solution and leading them to the assembly
site. Figure \ref{fig:assembly}b illustrates this process. We show
below that for a T=3 virus this mechanism can accelerate
self-assembly by approximately 15 times.

We consider a dilute solution of virus CPs with molecules of its
ss RNA genome. For the most of this paper we assume that
concentrations of the protein $c \sim 2Mc_{R}$, where $c_R$ is
the concentration of ss RNA and $M$ is the number of proteins in the
assembled virus (for T=3 viruses $M = 60{\rm T} = 180$). In this case there
are enough proteins in the system in order to assemble the virus
around each ss RNA molecule and $c$ changes weakly in the course
of assembly. Viruses, however, self-assemble only when the
concentration $c$ of CP is larger than some threshold
concentration $c_1$~\cite{Bruinsma1}, which is similar to the
critical micelle concentration for the self-assembly of
surfactant molecules~\cite{colloidal}. The critical concentration
$c_1$ can be estimated as
\be c_1 \approx \frac{1}{v} \exp[-(\epsilon_e+\epsilon_p)/k_BT],
\label{eq:first} \ee
where $v$ is the CP volume, $\epsilon_e$ is the absolute value of
the electrostatic adsorption energy of the CP N-terminal tail to
ss RNA, and $\epsilon_p$ is the absolute value of the CP-CP
attraction energy in the capsid (per CP). Both $\epsilon_e$ and
$\epsilon_p$ can be of the order of $10~k_{B}T$, so that the
critical concentration $c_1$ can be very small. In this paper we
always assume that $c \gg c_1$. As shown in Ref.~\cite{Hu}, in a
partially assembled capsid, CP sticks to a piece of ss RNA of the
length equal to the tail length $L$ (Fig. \ref{fig:assembly}a). A
partially assembled capsid with $m < M$ CPs encapsulates the length
$mL$ of ss RNA . To continue this process the next $(m + 1)$th CP
should attach itself to the partially assembled capsid at the
site, where ss RNA goes out of the capsid (see Fig.
\ref{fig:assembly}b) and this CP gets more nearest neighbors.
We call this slowly moving site "the assembly site". It has the
size of the order of the size $r$ of CP (see Fig.
\ref{fig:assembly}b).

\begin{figure}[htb]
\centering
\includegraphics[width=0.45 \textwidth]{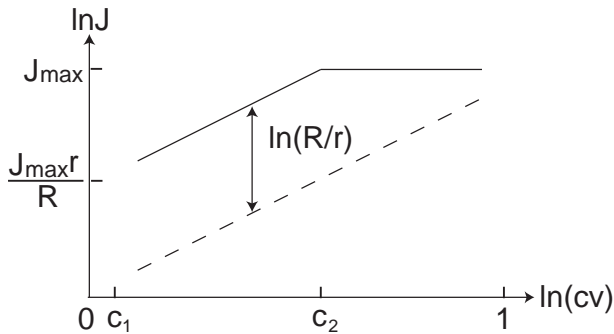}
\caption{Schematic plot of the diffusion limited self-assembly rate
$J$ as a function of the protein concentration $c$. The full line
is for the sliding of capsid proteins on ss RNA. The
rate for the slower three-dimensional diffusion is
shown by the dashed line.} \label{fig:rate}
\end{figure}

CPs diffuse to the assembly site through the bulk water. For $c
\gg c_1$ one can neglect the dissociation flux from the assembly
site. In this case the net rate of assembly (the number of CP
joining the capsid per unit time) is equal to the rate at which
diffusing CP find the absorbing sphere with the radius
$r$. It is equal to the Smoluchowski three-dimensional reaction rate~\cite{Smoluchowski}
\be J_3 = 4 \pi D_3 r c, \label{eq:Smol} \ee
where $D_3$ is the diffusion coefficient of CP in water. The rate
$J_3$ as a function of CP concentration $c$ is plotted in Fig.
\ref{fig:rate} by the dashed straight line.

Our main idea is that the long chain of yet unassembled ss RNA
outside of the capsid provides an additional route for the diffusion of
CPs to the assembly site, in analogy to the
well-known faster-than-diffusion locating of the specific site on DNA for
a protein~\cite{Riggs,BPJ}. The
dramatic enhancement of the assembly rate is achieved because, due
to the Boltzmann factor $\exp[\epsilon_{e}/k_BT]$, the three
dimensional concentration of CP on unassembled chain of ss RNA is
larger than the bulk concentration $c$. This concentration can be
estimated using the cylinder with cross-section $v^{2/3}$ build
around RNA as the axis: it is equal to the number of CPs per unit
length of ss RNA divided by $v^{2/3}$. At large distances the
one-dimensional flux of CP sliding on the ss RNA should
be balanced by the three dimensional diffusion flux of CP to the
ss RNA. This balance determines the radius $\xi$ of the sphere
around the assembly site at which two fluxes match each other and
the crossover between three-dimensional and one-dimensional
diffusions of CP takes place. The ss RNA coil inside this radius
is called antenna.

The maximum possible antenna size is the characteristic size $R
\sim (p\mathscr{L}_{e})^{1/2}$ of the unassembled portion of ss
RNA with length $\mathscr{L}_{e} = \mathscr{L} - mL$. (Here we
assume the ss RNA is a flexible Gaussian coil with the persistence
length $p\sim 2b \sim 1.5$~nm, where $b \simeq 0.7$~nm is the
monomer size, and do not account for the excluded volume
interaction.) In the case when $\xi = R$, the whole ss RNA
adsorbs CPs arriving by three-dimensional diffusion and provides a
path of fast one-dimensional diffusion to the assembly site (See
Fig. \ref{fig:assembly}). As a result, in this case the size $R$
replaces the protein size $r$ in Eq. (\ref{eq:Smol}) leading to a
much faster rate
\be J = 4 \pi D_3 R c, \label{eq:onedrate} \ee
which is shown in Fig. \ref{fig:rate} by the part of the solid line
parallel to the dashed one. Equation (\ref{eq:onedrate}) is correct
until CPs adsorbed on the unassembled chain of ss RNA are still sparse
and do not block each other's diffusion on ss RNA. Let us use the
notation $c_2$ for the concentration $c$, where the
antenna becomes saturated by CPs and the dependence of the
self-assembly rate $J$ on $c$ saturates roughly speaking at the
level $J_{max}=4\pi D_3 r/v$, which is the Smoluchowski rate $J_{3}$
at $c\sim 1/v$ (see the solid line in Fig. \ref{fig:rate}). It was
shown in Ref.~\cite{BPJ} that if $\xi \leq R$
\be c_{2} = \frac{1}{v} \exp[-\epsilon_e/k_BT]\approx c_{1}
\exp[\epsilon_p/k_BT]. \label{eq:second} \ee
We see that the largest enhancement $R/r$ of the self-assembly rate
$J$ can be achieved in the range of relatively small CP
concentrations $c_{1} \ll c \ll c_2$. For a typical T=3 virus the ss
RNA genome consists of 3000 bases, so that the length $\mathscr{L}
\sim 2100~$nm and $R\sim 60~$nm. Using $r \sim 4~$nm, we arrive at
the acceleration factor $R/r \sim 15$. One can calculate the assembly time
$\tau_a$ limited by diffusion. As we said above for $c \sim
2Mc_{R}$, the concentration of proteins $c$ can be regarded as a
constant. Thus, the assembly time with the help of antenna $\tau_a$ is given
by
\be \tau_a \approx \int_0^{M} \frac{dm}{4\pi cD_3[(M-m)Lp]^{1/2}} =
\frac{2M^{1/2}}{4\pi cD_3(Lp)^{1/2}}, \label{eq:time}\ee
while according to Eq. (\ref{eq:Smol}), the assembly time without antenna is simply $\tau_0 = M/(4\pi cD_3r)$. Since $(Lp)^{1/2} \sim
4$~nm, we can neglect the difference between $(Lp)^{1/2}$ and $r$,
and arrive at the assembly time with the help of antenna $M^{1/2}
\approx 14$ times shorter than $\tau_0$.

Strictly speaking, these estimates are correct only for
self-assembly with a homopolymeric ss RNA or a synthetic negative
polyelectrolyte~\cite{Brancroft}. For these cases, a small
additional acceleration by a factor 2 or 3 can be provided by the
excluded volume effect. On the other hand, the native ss RNA is more
compact than gaussian one due to hydrogen bonds forming hairpins and
thus the estimated acceleration rate can be reduced by a factor
between 2 or 3. Above we for simplicity replaced $\xi$ by its
maximum value $R$. The actual calculation of the antenna size $\xi$
can follow the logic of the scaling estimate for the search rate of
the specific site on DNA by a protein in Ref.~\cite{BPJ}. In our
case, the assembly site plays the role of the target site (diffusion
sink) for the protein, the unassembled chain of ss RNA plays the
role of DNA and the Coulomb attraction energy of N-terminal tails to
the unassembled ss RNA is analogous to the non-specific binding
energy of diffusing protein on DNA. One may argue that the virus
self-assembly problem is different, because ss RNA plays a dual
role. It is not only an antenna for the sliding CPs, but ss RNA
itself also moves to the assembly site, where it gets packed inside
the capsid (each newly assembled CP consumes the length $L$ of ss
RNA). However, for a small concentration $c$ in the range $c_{1} \ll
c \ll c_2$, where the unassembled ss RNA chain is weakly covered by
CPs, the velocity of ss RNA drift in the direction of assembly site
is much smaller than the average velocity of CP drift along ss RNA.
Thus, for the calculation of the assembly rate at a given length of
the unassembled ss RNA chain we can use the approximation of static ss RNA.
This brings us back to the problem of proteins
searching for the specific site on DNA~\cite{BPJ}. Note that this
means that the idea of self-assembly from the prepared linear ss
RNA-protein complex~\cite{Zlotnick04, Vogt95} is literally correct
only at $c > c_2$.

It is shown in Ref.~\cite{BPJ} that for a flexible ss RNA, the
antenna size reads $\xi \sim b(yd)^{1/3}$,
where $y = \exp(\epsilon_e/k_{B}T)$, $d = D_1/D_3$ and $D_1$ is
the one-dimensional diffusion coefficient of the protein sliding
on ss RNA. This result remains correct as long as the
antenna size $\xi$ is smaller than the ss RNA coil size
$R$. The energy  $\epsilon_e$ of adsorption of the N-terminal tail
with approximately 10 positive charges on ss RNA can be as large
as $10k_{B}T$. For $d=1$ we get $\xi \sim 30$ nm, while $R \sim
60$~nm. Thus, a simple estimate leads to the antenna length $\xi$
somewhat smaller than $R$.

There are, however, two reasons why $\xi$ may easily reach its
maximum value $R$. First, some viruses self-assemble from
dimers~\cite{Casini,Zlotnick04}. Naturally dimers with their two positive
tails bind to ss RNA with the twice larger energy $2\epsilon_e$.
This easily makes $\xi > R$. ii) The theory of Ref.~\cite{BPJ}
assumes that a sliding protein molecule has only one positive
patch, where it can be attached to a double helix DNA. Even if two
distant along the chain pieces of DNA come close in the
three-dimensional space, such protein can not simultaneously bind
both pieces and, therefore, can not crawl between them without
desorbing to water and losing the binding energy $-\epsilon_e$.
For a globular protein this is quite a natural assumption. On the
other hand, for CP attached to ss RNA by a flexible N-terminal
tail, the tail can easily cross over (crawl) between the two
adjacent pieces of the same ss RNA molecule losing only small
fraction of the energy $-\epsilon_e$.
This should lead to faster protein diffusion on ss RNA and may easily push $\xi$ up to $R$.

Let us discuss ideas of three in vitro experiments, which can verify
the role of ss RNA antenna in virus self-assembly. In the first
experiment, one breaks ss RNA molecule into $K\gg 1$ short pieces of
approximately equal length. It was shown~\cite{Rein,Vogt} that the
assembly is possible even when $K \sim M/2$, because in order to
glue CPs short ss RNA should bind two N-terminal tails of
neighboring proteins in the capsid. Virus assembly from short ss RNA
pieces goes consecutively through two different diffusion limited
stages. In the first stage, capsid fragments (CFs) made of $M/K$
proteins self-assemble on each short ss RNA molecule. According to
Eq.~\ref{eq:time}, the time necessary for this stage is proportional
to $(M/K)^{1/2}$ and is much shorter than the assembly time $\tau_a$
with the intact ss RNA. The second stage, where CFs aggregate to
form the whole capsid takes much larger time $\tau_{as}$ ($s$ stands
for short). In order to calculate $\tau_{as}$ we assume that when
two CFs with $n$ CPs each collide, they can relatively fast
rearrange their ss RNA and CPs in order to make one bigger CF with
$2n$ CPs. We also assume that at any time $t$ all CFs are
approximately of the same size $n(t)$. Then the concentration of
such CFs is $c(n) = c_{R}M/n(t)$, where $c_R$ is the concentration
of original intact ss RNA. Therefore, the time required for doubling
of a CF can be estimated from Eq.~(\ref{eq:Smol})
\be \tau(n) = \frac{1}{4\pi D_{3}(n) r(n) c(n)} = \frac{n}{4\pi
D_{3}(n) r(n) c_R M}, \label{eq:taun} \ee
where $D_{3}(n)$ and $r(n)$ are diffusion coefficient and effective
radius of a CF with $n$ CPs. Since the diffusion coefficient is
inversely proportional to the droplet radius, the product $D_{3}(n)
r(n) = k_B T/6\pi \eta$, (where $\eta$ is the water viscosity), is the
same constant as $ D_{3} r $ for a single protein. One collision of
droplets transfers $n$ CPs to the growing CF. Therefore, the average
time needed to add one CP to the growing CF $ \tau_1 = \tau(n)/n =
1/4\pi D_{3} r M c_{R} $ does not depend on $n$. In other words, the
number $n(t)$ of CP per CF increases at a constant rate. The
assembly ends when $n$ reaches $M$. Therefore, the assembly time is
given by
\be \tau_{as} \simeq M \tau_1
\simeq \frac{1}{4\pi c_R D_3 r}. \label{eq:tauas} \ee
Above equation shows the assembly time depends on $Mc_R$ which stands for
the concentration of CP involved in the CF aggregation.
However $\tau_{as}$ has no dependence on $K$.
Comparing Eqs. \ref{eq:time} and \ref{eq:tauas}, we obtain that at
$c \sim 2Mc_{R}$
\be \frac{\tau_{as}}{\tau_a} \sim M^{1/2} \frac{(Lp)^{1/2}}{r} \sim
M^{1/2}\gg 1. \ee
We see that the virus assembly time with short ss RNA pieces is
much larger than that for the intact ss RNA. This happens due to
the breaking of big antenna of the original ss RNA.
\begin{figure}[htb]
\centering
\includegraphics[width=0.45 \textwidth]{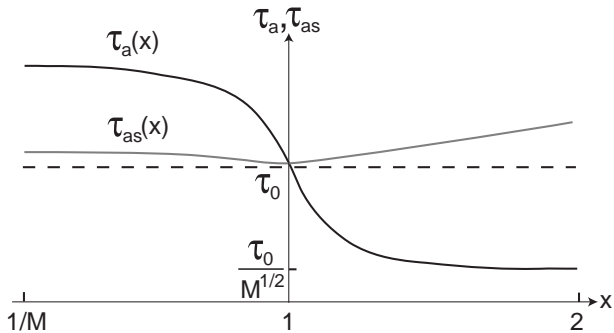}
\caption{Self-assembly times plotted schematically as a function of
$x = c/Mc_R$.
$\tau_0 = M/(4\pi cD_3r)$ is the assembly time without the effect of ss RNA at $x > 1$.
The dark and gray lines correspond to intact ss RNA and short RNA pieces respectively.}
\label{fig:kinetictrap}
\end{figure}

In the second experiment, we return to the intact ss RNA and
discuss what happens when we vary relative concentrations of CP
and ss RNA $x=c/Mc_{R}$, for example, keeping $c = const$ and
changing $c_R$. Until now we assumed that $x \sim 2$, i.e. we have
marginally more proteins than it is necessary to assemble a virus
at every ss RNA. If $x \gg 1$ the assembly time $\tau_a$ is
practically the same as that at $x \sim 2$ and is given by Eq.
\ref{eq:time}. Let us now consider much larger $c_R$, for which $x
\ll 1$. Here situation changes dramatically. There are two
assembly stages. In the first stage, a CF is assembled with part
of each ss RNA molecule, leaving the rest of the ss RNA molecule
as a tail. This assembly uses up all the proteins and stops, when
all CFs are still much smaller than the complete capsid and their
ss RNA tails are long (see, for example, Fig.
\ref{fig:assembly}b). This state is essentially a kinetic trap. If
energies $\epsilon_e$ and $\epsilon_p$ are much larger than
$k_BT$, CFs on different ss RNA molecules can not exchange CPs
trough the solution or via collision of their ss RNA tails. They
can grow only via CF-CF collisions, while merging on one ss RNA and releasing
the other empty one. We explained above, at $x>1$ (CPs are in
excess), CFs without RNA tails produce a capsid during
time given by Eq. \ref{eq:tauas}. On the other hand, at $x < 1$,
only occupied by CP ss RNA molecules take part in the aggregation
and in order to get the assembly time, $c_R$ in Eq. \ref{eq:tauas}
should be replaced by $c/M$, which does not depend on $x$. However, due to the
long ss RNA tail, a CF diffuses slower than it does without a
tail. The time $\tau_a(x)$ grows substantially with decreasing
$x$, because with more ss RNA, the initial CFs have fewer CPs and
longer ss RNA tails. This time saturates at $x \sim 1/M$, where
$c=c_R$ and each CF has only one protein and the longest ss RNA
tail. Thus, a long antenna accelerates assembly at $x
>1$ and decelerates it at $x <1 $. This behavior of $\tau_{a}(x)$
is schematically plotted in Fig. \ref{fig:kinetictrap}.

In the third experiment we can combine the first two and break ss
RNA into pieces at several different values of $x$. At $x < 1$ a
CF gets a shorter tail of ss RNA and larger mobility, so that
assembly is faster than for intact ss RNA. When $x > 1$, the
assembly time grows according to Eq.~(\ref{eq:tauas}) with
decreasing $c_R$ (increasing $x$). This is because the smaller the
ss RNA concentration, the harder for the CFs to collide with each
other and form larger CFs. In other words, kinetics is determined
only by CPs already assembled in CFs and their number decreases
with growing $x$. We illustrate such nontrivial role of broken ss
RNA in Fig. \ref{fig:kinetictrap}.

Now let us give some numerical estimates for $c_1$, $c_2$ and $\tau_0$ for the in vitro
assembly. Using the radius of CP $r \sim 4$ nm, we obtain $c_1
\sim 0.1$ nM and $c_2 \sim 1$ $\mu$M from Eqs.~(\ref{eq:first}) and (\ref{eq:second}).
For $c \sim 1$ nM and the diffusion coefficient $D_3 \sim 2\times10^{-7} {\rm cm}^2/{\rm s}$, the
assembly time $\tau_0$ is about 10 min. At excess of CP, ss RNA antenna
reduces it to $\tau_a \sim 1$ min. At excess of ss RNA roughly
speaking $\tau_a$ increases to $2\tau_0$. One can make $\tau_a$ even larger
using much longer than native ss RNA.

In conclusion, we studied the role played by unassembled tail of
ss RNA, which we call antenna. We showed that one-dimensional
diffusion accelerates the virus self-assembly more than ten times
when proteins are in excess with respect to RNA. On the other hand
when RNA is in excess long tail of ss RNA slows down the assembly.
We discussed several experiments which can verify the role of
antenna. Although in this paper we focus on viruses for which CPs
have long positive N-terminal tails, our idea can be also applied
to the case where a CP binds to ss RNA by its positive patch. Our
ideas are applicable beyond icosahedral viruses, for example, to
the assembly of immature retro-viruses such as RSV or
HIV~\cite{Rein,Vogt95,Vogt}.

We are grateful to V. A. Belyi, R. Bruinsma, A. Yu. Grosberg, T. T.
Nguyen, A. Rein, Iu. Rouzina, V. M. Vogt and A. Zlotnick for useful
discussions.


\end{document}